\pdfoutput = 1
\documentclass[%
 reprint,
 amsmath,amssymb,
 aps,
]{revtex4-2}

\usepackage{graphicx}
\usepackage{dcolumn}
\usepackage{bm}
\usepackage{braket}
\usepackage{natbib}

\begin{document}

\preprint{APS/123-QED}

\title{Parametric t-Stochastic Neighbor Embedding With Quantum Neural Network}

\author{Yoshiaki Kawase}
\affiliation{%
Graduate School of Engineering Science, Osaka University\\
1-3 Machikaneyama, Toyonaka, Osaka 560-831, Japan
}%
\email{yoshiaki.kawase@qc.ee.es.osaka-u.ac.jp}
\author{Kosuke Mitarai}%
 \email{mitarai@qc.ee.es.osaka-u.ac.jp}
\affiliation{%
Graduate School of Engineering Science, Osaka University\\
1-3 Machikaneyama, Toyonaka, Osaka 560-831, Japan
}
\affiliation{%
Center for Quantum Information and Quantum Biology, Osaka University, Japan.
}%
\affiliation{%
JST, PRESTO, 4-1-8 Honcho, Kawaguchi, Saitama 332-0012, Japan.
}

\author{Keisuke Fujii}
\email{fujii@qc.ee.es.osaka-u.ac.jp} 
\affiliation{%
Graduate School of Engineering Science, Osaka University\\
1-3 Machikaneyama, Toyonaka, Osaka 560-831, Japan
}
\affiliation{
Center for Quantum Information and Quantum Biology, Osaka University, Japan.
}%
\affiliation{
RIKEN Center for Quantum Computing, Wako Saitama 351-0198, Japan.
}

\date{\today}

\begin{abstract}
t-Stochastic Neighbor Embedding (t-SNE) is a non-parametric data visualization method in classical machine learning. 
It maps the data from the high-dimensional space into a low-dimensional space, especially a two-dimensional plane, while maintaining the relationship, or similarities, between the surrounding points. 
In t-SNE, the initial position of the low-dimensional data is randomly determined, and the visualization is achieved by moving the low-dimensional data to minimize a cost function. 
Its variant called parametric t-SNE uses neural networks for this mapping.
In this paper, we propose to use quantum neural networks for parametric t-SNE to reflect the characteristics of high-dimensional quantum data on low-dimensional data. 
We use fidelity-based metrics instead of Euclidean distance in calculating high-dimensional data similarity. 
We visualize both classical (Iris dataset) and quantum (time-depending Hamiltonian dynamics) data for classification tasks. 
Since this method allows us to represent a quantum dataset in a higher dimensional Hilbert space by a quantum dataset in a lower dimension while keeping their similarity, 
the proposed method can also be used to compress quantum data for further quantum machine learning.
\end{abstract}

\maketitle


\section{\label{sec:introduction1}Introduction}

Visualization of high-dimensional data is an important subfield of machine learning.
It allows us to intuitively interpret the data and understand possible patterns in them.
As visualization often involves mapping of the original data to a low, typically two or three, dimensional space, the techniques for visualizations are also useful for compression of data or preprocessing before applying another machine learning techniques.
Prototypical examples of visualization techniques include t-stochastic neighbour embedding (t-SNE) \cite{van2008visualizing}.

Such techniques have been proved to be useful also for machine learning of quantum states.
Ref. \cite{PhysRevE.97.013306} applied various visualization methods to detect quantum phase transitions in the Hubbard model, where they generated states by quantum Monte Carlo simulations.
They have found that t-SNE is the most promising technique in this type of the application.
Ref. \cite{Yuan2021} applied t-SNE to visualize quantum states represented as a matrix product states.
They successfully visualized quantum phase transitions in spin models such as transverse field Ising model.

However, the application of machine learning techniques running on classical computers for this purpose is intrinsically limited to the case where the target quantum states have efficient classical representations.
The use of quantum computers can provide advantage in widening the range of quantum states that can be used as inputs.
In fact, in Ref.~\cite{Okada2022}, identification of topological quantum phase has been proposed by using a clustering algorithm using a quantum computer.
This approach is motivated by the fact that there is a family of quantum states that are useful for machine learning of classical data but cannot be efficiently represented by classical computers \cite{liu2021}.

Here, we propose to use quantum neural networks (QNNs), which use parametric quantum circuit to construct a machine learning model, for visualization of quantum states.
Our proposal is based on parametric t-SNE \cite{van2009learning}, which is a visualization technique where we employ neural networks to map high-dimensional data into low-dimensional space.
The mapping is optimized so as to keep the similarity of the data points, which is defined from the distance in the respective spaces, unchanged.
Our idea is to use QNNs instead of the classical neural network.
This allows us to directly use quantum states as inputs, which may be useful for studying complex quantum systems and certain machine learning problems that are hard classically.
The similarity of the quantum states can be defined from fildelity(-like) measures.
Our method optimizes the parameters in a QNN so that the respective quantum states are mapped to low-dimensional points, which are defined as expectation values of certain observables at the output of the QNN, while maitaining the similarity among the points.

We also conduct numerical verification of our proposed method. 
First, we use the Iris flower dataset \cite{Dua:2019} to test if it can be successfully applied to classical data. 
Second, we visualize the quantum states time-evolved under the transverse-field Ising Hamiltonian.
In the visualization of quantum data, 
we could not effectively generate a visualization with distinct clusters when using default cost function usually utilized for parametric t-SNE.
To deal with this problem, we introduce a hyper-parameter into the cost function to adapt the scaling of the low-dimensional data.
With these successful demonstrations, we believe that the proposed method would be a powerful tool to visualize and analyse both classical and quantum datasets.

This paper is organized as follows. 
In section \ref{sec:background}, we briefly review the t-SNE. 
In section \ref{sec:proposed_method}, we describe our proposed method. 
In section \ref{sec:numerical_experiments}, we perform numerical experiments to verify our proposed method. 
In section \ref{sec:conclusion}, we describe the conclusion and future work.

\section{\label{sec:background}Background}
\subsection{\label{parametric_tsne}t-Stochastic Neighbor Embedding}
The t-Stochastic Neighbor Embedding (t-SNE) \cite{van2008visualizing}, also called non-parametric t-SNE,  is a classical machine learning method for visualizing high-dimensional data.
The idea of t-SNE is to map data points in the original high-dimensional space to points in a low-dimensional space while keeping the similarity among the points.
The map is determined by minimizing the KL-divergence between the similarity of data distributions in the high- and low- dimensional space. 

In detail, the t-SNE defines the similarity  between a high-dimensional datapoint $x_i$ and another datapoint $x_j$ by the following joint probability \cite{van2008visualizing}.
\begin{eqnarray} \label{eq:p_similarity}
p_{ij} = \frac{p_{i|j}+p_{j|i}}{2N}, 
\end{eqnarray}
where,
\begin{eqnarray} \label{eq:p_conditional_prob}
p_{j|i} &=& \frac{ \exp{ \left( -\| x_i-x_j \|^2 /2 \sigma_i^2 \right) }}{\sum_{k \neq i} \exp{ \left(-\| x_i-x_k \|^2 /2 \sigma_i^2 \right) } }, \\
p_{ii} &=& 0, \nonumber
\end{eqnarray}
and $\sigma_i$ are parameters determined from the following quantity called perplexity of data $x_i$.
\begin{eqnarray}
Perp_i=2^{-\sum_j p_{j|i} \log_2{p_{j|i}}}. \nonumber
\end{eqnarray}
The value of $\sigma_i$ is set so as to make $Perp_i$ a user-specified value (typically between $5$ and $50$). 
The similarity between the low-dimensional data points $y_i$ and $y_j$ is defined by the following equation using Student t-distribution with one degree of freedom \cite{van2008visualizing}. 
\begin{eqnarray}
q_{ij} &=& \frac{ \left( 1+\| y_i - y_j \|^2 \right)^{-1}}{ \sum_{k} \sum_{l \neq k} \left( 1+\| y_k-y_l \|^2 \right)^{-1}}, \nonumber \\
q_{ii} &=& 0. \nonumber
\end{eqnarray}
The t-SNE determines a low-dimensional point $y_i$ corresponding to a datapoint $x_i$ by iteratively minimizing the cost function $C(\{y_i\})$ defined as the Kullback-Leibler (KL) divergence between a joint probability distribution in the high- and low- dimensional data 
\begin{eqnarray} \label{eqn:cost_function_tsne}
C(\{y_i\}) = \sum_i \sum_j p_{ij} \log{ \frac{p_{ij}}{q_{ij}} }.
\end{eqnarray}
The gradient of the cost function for $y_i$ is described as 
\begin{eqnarray} \label{eqn:grad_cost_function_tsne}
\frac{\partial C}{\partial y_i} = 
4 \sum_j (p_{ij}-q_{ij})(y_i-y_j)(1+\| y_i - y_j \|^2)^{-1}.
\end{eqnarray}
In optimizing the cost function, each $y_i$ is initially placed in a random position and moved to minimize the cost function.

\subsection{Parametric t-SNE}
Parametric t-SNE \cite{van2009learning} is a variant of t-SNE which discards the direct optimization of the low-dimensional points $\{y_i\}$ but uses neural network to map $x_i$ to $y_i$.
In this method, each low-dimensional point $y_i$ is generated by $y_i = f(x_i|\theta)$, where $f(x_i|\theta)$ is an output of the neural network with an input $x_i$ and network weight $\theta$.
We optimize $\theta$ to minimize the cost function Eq.~(\ref{eqn:cost_function_tsne}).
An important distinction between the t-SNE and the parametric t-SNE is that the latter can easily generate a low-dimensional point for a new input as we explicitly construct the mapping from $x_i$ to $y_i$.

\subsection{Application of t-SNE for quantum systems}
Recently, t-SNE has also been applied to the field of quantum physics by Yuan \textit{et al.}
\cite{Yuan2021}.
They considered the visualization of quantum phase transitions by applying t-SNE to (approximate) ground states $\ket{\psi_i}$ of certain Hamiltonians $H_i$.
Their approach is to use,
\begin{eqnarray} 
p_{j|i} = \frac{ \exp{ \left( - d(\psi_i,\psi_j)^2 /2 \sigma_i^2 \right) }}{\sum_{k \neq i} \exp{ \left(- d(\psi_i,\psi_k)^2 /2 \sigma_i^2 \right) } },
\end{eqnarray}
instead of Eq. (\ref{eq:p_conditional_prob}), where the distance $d(\psi_i,\psi_j)$ is defined by the negative logarithmic fidelity
\begin{eqnarray}
d(\psi_i,\psi_j) = -\log\left(\left|\braket{\psi_i|\psi_j}\right|\right).
\end{eqnarray}
It has been shown that this approach can successfully visualize and identify quantum phase transitions of one-dimensional spin chains.
In Ref. \cite{Yuan2021}, they also considered the visualization of classical data by using the so-called ``quantum feature map''
via t-SNE.
We share the idea with Ref. \cite{Yuan2021} in the sense of visualization of quantum states via t-SNE.
However, our approach differs essentially from Ref. \cite{Yuan2021}
in that a parameterized quantum model is employed as low dimensional data of parameteric t-SNE.

\section{\label{sec:proposed_method}Parametric \lowercase{t}-SNE with quantum circuits}
\begin{figure*}
    \centering
    \includegraphics[clip, keepaspectratio, scale=0.5]{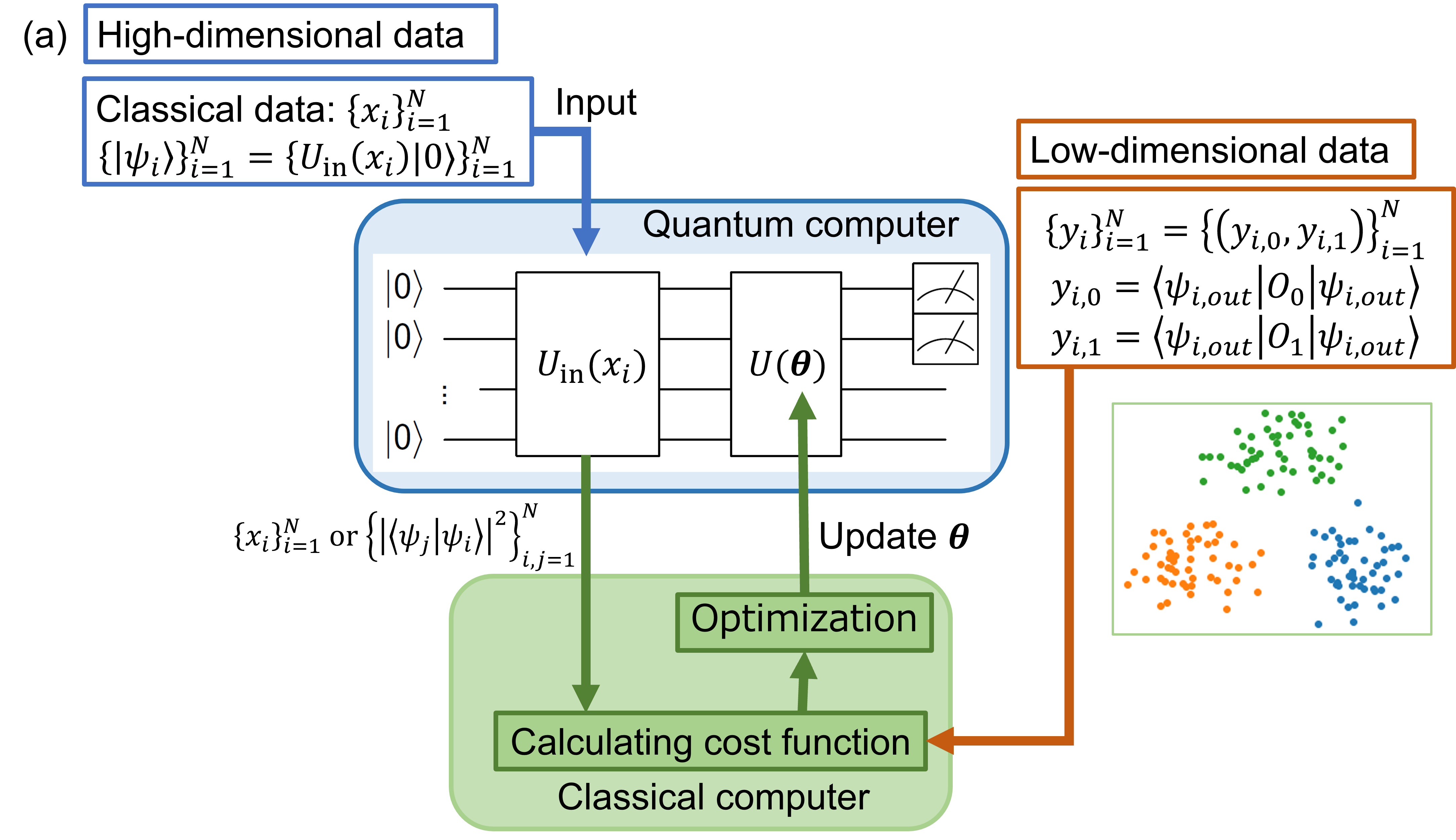}
    \includegraphics[clip, keepaspectratio, scale=0.5]{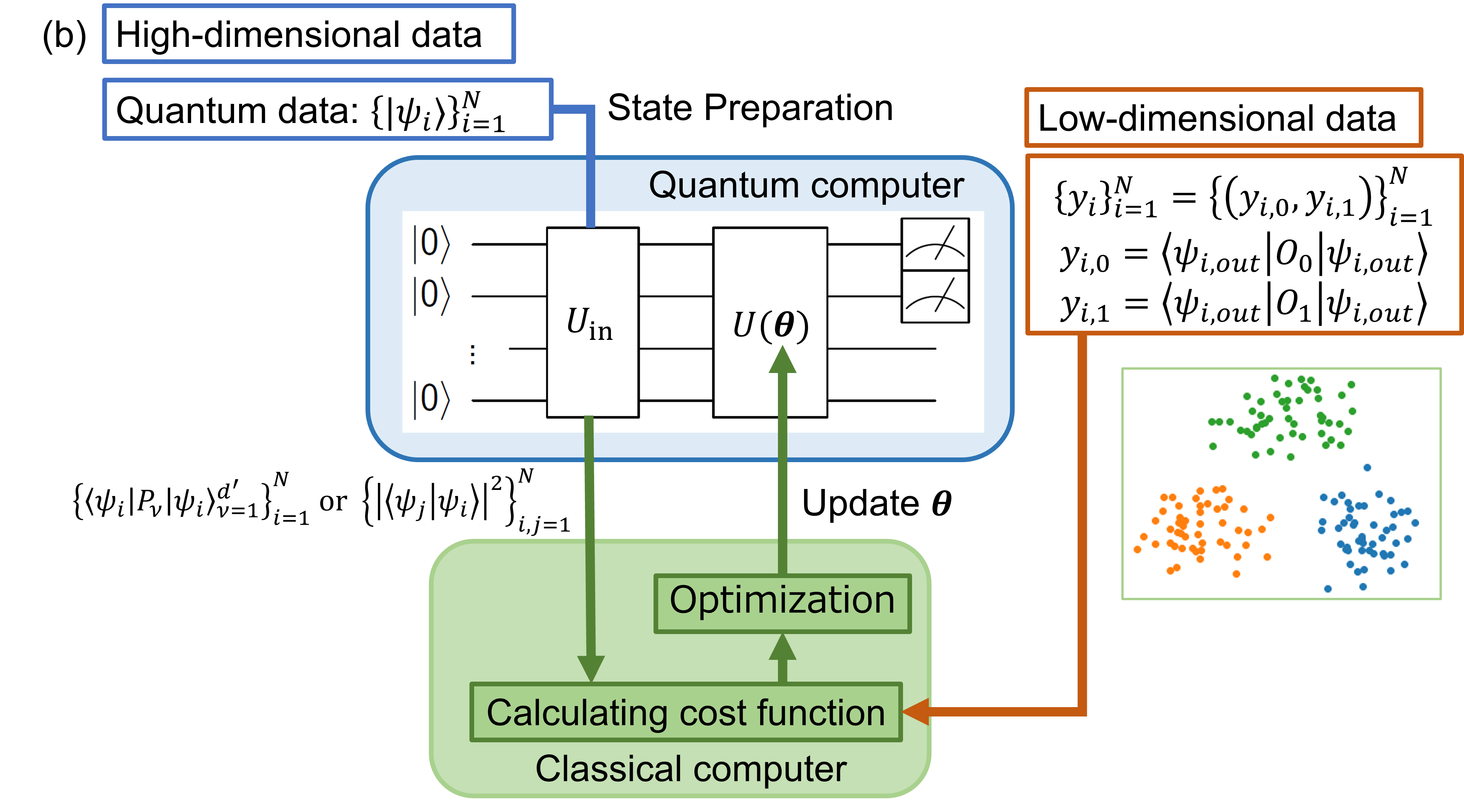}
    \caption{
    Figures (a) and (b) show the procedure for visualizing classical and quantum data, respectively.
    We visualize classical and quantum data by mapping the high-dimensional data to the low-dimensional one. 
    The mapping is performed by the parametrized quantum circuit trained by minimizing a cost function so as to maintain the similarity of surrounding points. 
    The similarity of high-dimensional data is defined by either expectation values, fidelities, or classical data. 
    The similarity of low-dimensional data is determined by expectation values, and we plot them in a two-dimensional plane.
    }
    \label{fig:proposed_method}
\end{figure*}

In this work, we propose using quantum circuits to construct the parametric model $f(x|\theta)$ to generate low-dimensional data.
The procedure of our proposed method for visualizing classical data is shown in Fig.~\ref{fig:proposed_method} (a).
More concretely, using a parameterized unitary circuit $U(x,\theta)$ that depends on both of a $d$-dimensional input $x$ and trainable parameters $\theta$, we generate a $d'$-dimensional datapoint $y$ by expectation values of $d'$ observables $\{O_\mu\}_{i=1}^{d'}$.
Then, we minimize the cost function defined in Eq.~(\ref{eqn:cost_function_tsne}) by optimizing the parameters $\theta$.
The concrete algorithm for classical inputs is as follows:
\begin{enumerate}
    \item Compute $p_{ij}$ for all pairs from $\{x_i\}_{i=1}^N$.
    \item For all $\{x_i\}_{i=1}^N$ and $\{O_\mu\}_{\mu=1}^{d'}$, evaluate $y_{\mu}(x_i, \theta) = \bra{0}U^\dagger(x_i,\theta) O_\mu U(x_i,\theta)\ket{0}$ using a quantum computer.
    \item Compute $q_{ij}$ for all pairs from $y_{\mu}(x_i, \theta)$ on a classical computer.
    \item Compute $C(\{y_{\mu}(x_i, \theta)\}_{i=1}^N)$ on a classical computer.
    \item Update $\theta$ so that $C$ is minimized.
\end{enumerate}
After the convergence, we expect the above protocol to find a quantum circuit $U(x,\theta)$ which can map $x_i$ to a low-dimensional space while preserving the similarity based on distances among the data.

An interesting extension of the above protocol is to take a quantum dataset consisting of quantum states $\{\ket{\psi_i}\}_{i=1}^N$ as the input data.
We depict the protocol in Fig.~\ref{fig:proposed_method} (b).
Here we can think of several ways to define the similarity of quantum data in a high dimensional space.
One possible way is to generate high-dimensional classical data by measuring a set of observables $\{ P_\nu \}_{\nu=1}^{d}$. Then their expectation values 
\begin{eqnarray}
\{ \langle  P_\nu \rangle \}_{\nu=1}^{d} 
\end{eqnarray}
are
used as the high-dimensional data.
Another possibility is to 
use a distance function of two quantum states defined via fidelity
\begin{eqnarray}\label{eq:infidelity-distance}
d(|\psi_i\rangle ,|\psi_k\rangle) = \sqrt{1-\left|\braket{\psi_i|\psi_k}\right|^2}.
\end{eqnarray}
Then similarity of two quantum states are defined as follows:
\begin{eqnarray} \label{pji_dist}
p_{j|i} = \frac{ \exp{ \left( - d(|\psi_i\rangle ,|\psi_j\rangle)^2 /2 \sigma_i^2 \right) }}{\sum_{k \neq i} \exp{ \left(- d(|\psi_i\rangle ,|\psi_k \rangle)^2 /2 \sigma_i^2 \right) } }.
\end{eqnarray}
This choice enables us to readily calculate $d(\psi_i,\psi_k)$ on a quantum computer by using standard techniques for overlap measurement such as swap test \cite{Buhrman2001}.

On the other hand, for the low-dimensional space,
we can use $U(\theta)$ that only depends on the trainable parameters $\theta$ at the step 2 in the classical input case since we do not have classical input $x$.
For the cost function $C$, we can use the same formulation as above, that is, we measure $y_{\mu}(\psi_i, \theta) =  \bra{\psi_i}U^\dagger(\theta) O_\mu U(\theta)\ket{\psi_i}$ for a certain set of observables $\{O_\mu\}$ and compute $C(\{y_{\mu}(\psi_i,\theta)\}_{i=1}^N)$. 

Note that if we employ quantum feature map,
\begin{eqnarray}
|\psi _i \rangle = U_{\rm in} (x_i) |0\rangle,
\end{eqnarray}
the similarity of classical input data $\{ x_i \}$
in high dimensional space
can also be calculated via the fidelities of the quantum states.

The possible advantage of the proposed method solely depends on the fact that there are quantum circuits that are hard to simulate classically \cite{arute2019quantum,aaronson2010computational,Bremner2010IQP}.
This means that we may be able to construct the map $f(x|\theta)$ which cannot be expressed by neural networks.
In fact, it is proven that, when using a very specific dataset, there is a machine learning task with rigorous quantum advantage \cite{liu2021}.
However, usefulness of quantum circuits for modeling a practical classical data are, in general, still in question.
Since one of our approaches assumes that the high-dimensional data is quantum data, we expect that the quantum model has an advantage in reproducing its similarity in low dimensions.
We leave such a research direction as a important future direction to explore.

\section{\label{sec:numerical_experiments}Numerical Experiments}
Here, we perform the numerical simulations of the proposed method.
First, let us describe the tools used in the experiments.
We use qulacs \cite{suzuki2021qulacs} to simulate a quantum circuit. 
We make a python wrapper to work with PyTorch\cite{NEURIPS2019_9015} 
to use the loss function and optimizer.
Our optimization is performed by Sharpness-Aware Minimization (SAM) \cite{foret2020sharpness,davda54_2020sam} 
with Adam\cite{kingma2014adam} as the base optimizer.
The overview of SAM is to consider the cost added the l2-regularization term, 
find the gradient at the point where that cost is the highest in the neighborhood, 
and descend from the current point according to that gradient.

\subsection{\label{subsec:vis_classical_data}Visualizing Classical Data}
\subsubsection{\label{subsubsec:app_proposed_method}Application of parametric t-SNE with quantum circuits}
\begin{figure*}
	\includegraphics[clip, keepaspectratio, scale=0.6 ]{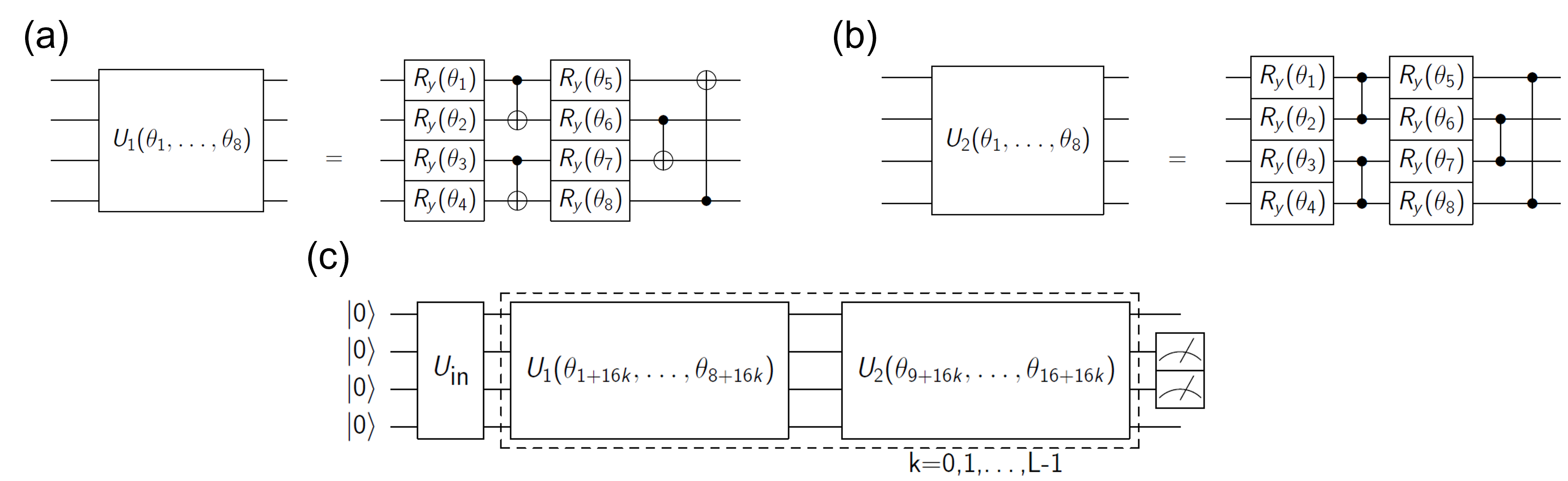}
	\caption{
	The quantum circuit to visualize data: 
	We input data by $U_{\mbox{in}}$, and alternately act $U_1$ and $U_2$ on the quantum state. The quantum circuit $U_1$ and $U_2$ is shown in (a) and (b), respectively.
	}
	\label{fig:iris_circuit}
\end{figure*}

We visualize the Iris flower dataset \cite{Dua:2019} with our proposed method. 
The dataset contains three classes and consists of four features. 
We normalize each feature between $-1$ and $+1$. 
The construction of our ansatz $U(x,\theta)$ and observables $\{O_\mu\}$ for the low dimensional space are as follows.
Let us define $R_y(\theta)=e^{i\theta \hat{Y}/2}$ and 
\begin{eqnarray} \label{eqn:input_cl}
U_{\mathrm{in}}(x_i) = R_y(x_{i,1}) \otimes R_y(x_{i,2}) \otimes \ldots \otimes R_y(x_{i,d}), \nonumber
\end{eqnarray}
where $x_{i,j}$ denotes the $j$th element of $i$th data $x_i$.
Also, let us define $U_1$ and $U_2$ as shown in Fig. \ref{fig:iris_circuit}.
We construct $U(x,\theta)$ as,
\begin{equation}
    U(x,\theta) = \prod_{k=0}^{L-1} \left[U_2(\theta_{9+16k:16+16k}) U_1(\theta_{1+16k:8+16k})\right] U_{\mathrm{in}}(x),
\end{equation}
where $\theta_{i:j}$ denotes $(j-i+1)$ dimensional vector containing $i$th to $j$th elements of $\theta$.
Specifically, we set the circuit depth $L=4$.
As for the output, we set $d'=2$ and $\{O_\mu\} = \{X_2, X_3\}$ to visualize the Iris dataset on two-dimensional plane.

\begin{figure*}
    \centering
    \includegraphics[clip, keepaspectratio, scale=0.50]{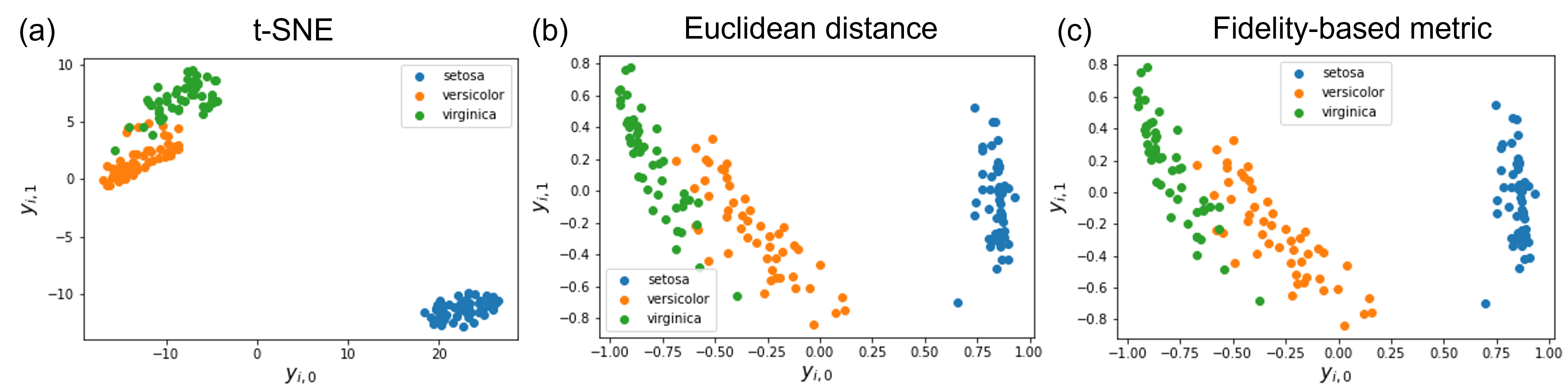}
    \caption{
    The figure (a) shows the visualization of the Iris flower dataset 
    by the classical machine learning method of t-SNE, 
    the figure (b) by the proposed method with the similarity based on Euclidean distance between high-dimensional data, 
    and the figure (c) by the proposed method 
    with the similarity based on the fidelity based metric between high-dimensional data.
    }
    \label{fig:iris_output}
\end{figure*}

We perform the simulation under the above settings and plot the results in Fig. \ref{fig:iris_output}. 
This figure shows that the data is clustered for each type of Iris flower.
The left figure (a) shows the visualization by the classical machine learning method of t-SNE, 
the middle figure (b) by the proposed method with similarity based on Euclidean distance.
In both two cases, the data are clustered by Iris flower varieties.

\subsubsection{\label{subsubsec:vis_iris_infidelity}Visualization with infidelity distance measure} 
Next, we show that the proposed method can be correctly visualized by calculating the similarity of quantum states. To this end, we will use, as a test case, a set of quantum states generated by a quantum feature map from the Iris dataset, before running the method on a truly physically meaningful quantum state.
We perform the parametric t-SNE using quantum features $\ket{\psi_i}$ defined as
\begin{equation}
    \ket{\psi_i} = U_{\mathrm{in}}(x_i)\ket{0},
\end{equation}
and the cost function associated with the distance measure defined in Eq. (\ref{eq:infidelity-distance}).
We show the result in Fig. \ref{fig:iris_output} (c), which implies that this protocol also work.

\subsection{Visualization of Quantum Data}
\subsubsection{Visualization based on Observables}
In this section, we perform the visualization of quantum states. 
As an example, let us consider the time-dependent two-body transverse field Ising model, which is employed in quantum annealing \cite{kadowaki1998quantum}.
We prepare quantum states time-evolved 
under the following Hamiltonian.
\begin{eqnarray}
    H = (1- t / \tau) \sum_i h_i X_i + ( t / \tau) \sum_{i<j} J_{ij} Z_i Z_j,
\end{eqnarray}
where $\tau$ denotes the total simulation time of the Hamiltonian. 
We perform the simulation under the Trotter decomposition\cite{trotter1959product,suzuki1976generalized}.
We set the number of qubits as four, $\tau=40$, time step $\Delta t=0.01$,
$h_i=1$, and $J_{ij}$ as uniform random number between $-1$ and $0$ or $0$ and $+1$.
We prepare $100$ samples each with positive or negative $J_{ij}$.

\begin{figure*}
    \centering
    \includegraphics[clip, keepaspectratio, scale=0.5]{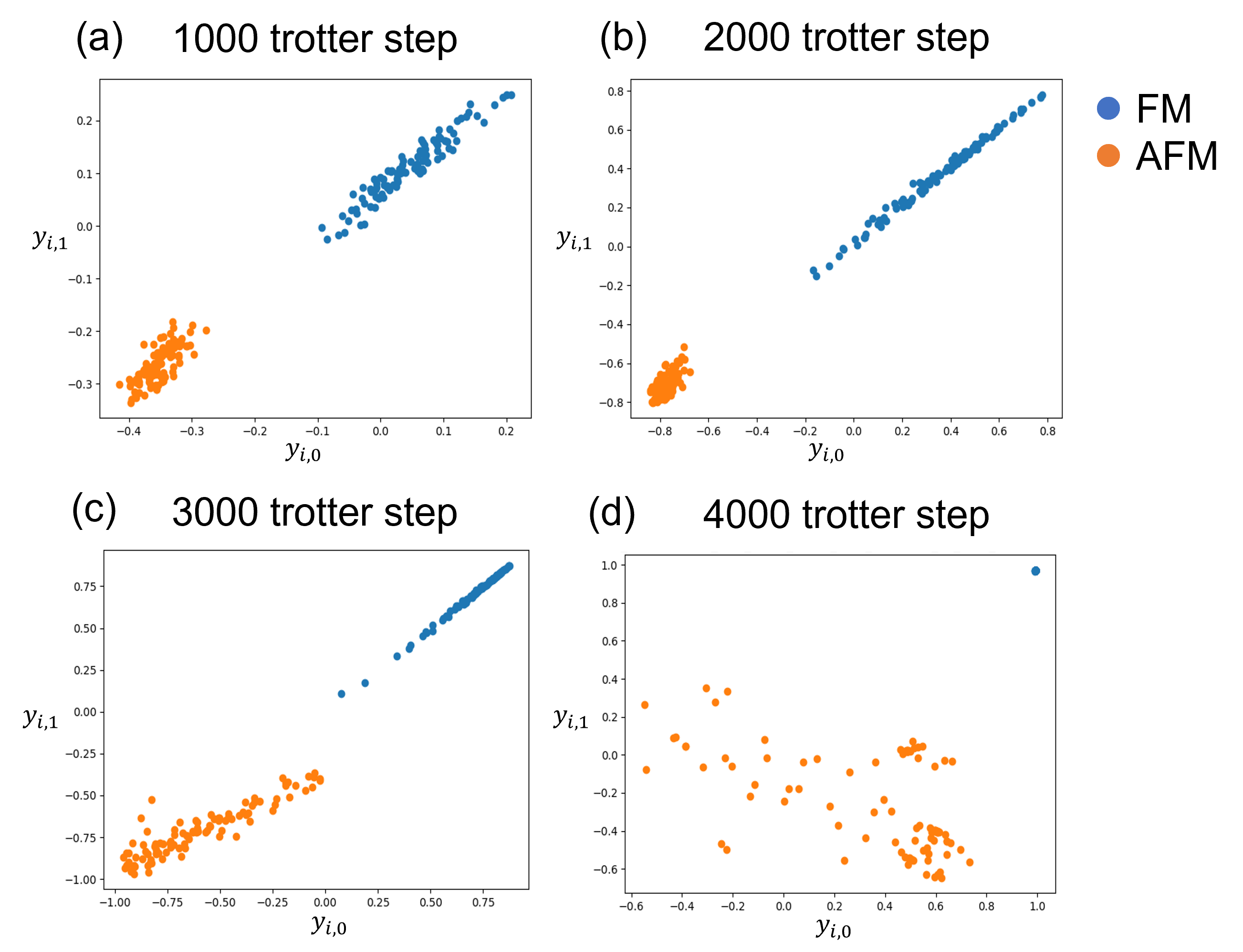}
    \caption{
    These figures illustrate the process of a time-dependent transverse-field two-body Ising model 
    when the calculation of the similarity in high-dimensional data is based on the observables.
    The initial state is $|+\rangle^{\otimes 4}$.
    The figure (a), (b), (c), and (d) visualize the quantum states of the $1000$th, $2000$th, $3000$th, and $4000$th trotter steps, respectively. 
    The blue and orange points correspond to the cases where $J_{ij}$ is positive and negative, respectively.
    }
    \label{fig:qa_output}
\end{figure*}

As explained before, we consider the two methods to compute 
similarity of the quantum states in high-dimensional space.
The first method is to consider the expectation values of input quantum states as high-dimensional data 
\begin{eqnarray}
x_i = 
(\langle \psi_{i,\mathrm{in}}|X_1|\psi_{i,\mathrm{in}} \rangle, \langle \psi_{i,\mathrm{in}}|X_2|\psi_{i,\mathrm{in}} \rangle, \ldots,
\langle \psi_{i,\mathrm{in}}|X_{n}|\psi_{i,\mathrm{in}} \rangle ), \nonumber
\end{eqnarray}
and calculate the similarity between the high-dimensional data points as we explained in Sec.~\ref{parametric_tsne}.
The low-dimensional data $\{y_i\}$ is defined similarly by the following equation using a certain constant value $a$:
\begin{eqnarray} \label{eq:low-dim}
y_i = a (\langle \psi_{i,\mathrm{out}}|X_2|\psi_{i,\mathrm{out}} \rangle, \langle \psi_{i,\mathrm{out}}|X_3|\psi_{i,\mathrm{out}} \rangle) .
\end{eqnarray}
The constant value $a$ is a hyper-parameter to adjust the scale of low-dimensional data.
We set the parameters $a=1$ and circuit depth $L=8$.
In this numerical experiment, 
we visualize the quantum states for every $1000$ trotter steps.
The result is shown in Fig.~\ref{fig:qa_output}.
In this figure, we can see that two clusters corresponding to the sign of the coupling constants are formed.

\subsubsection{Visualization with infidelity metric}
\begin{figure*}
    \centering
    \includegraphics[clip, keepaspectratio, scale=0.5]{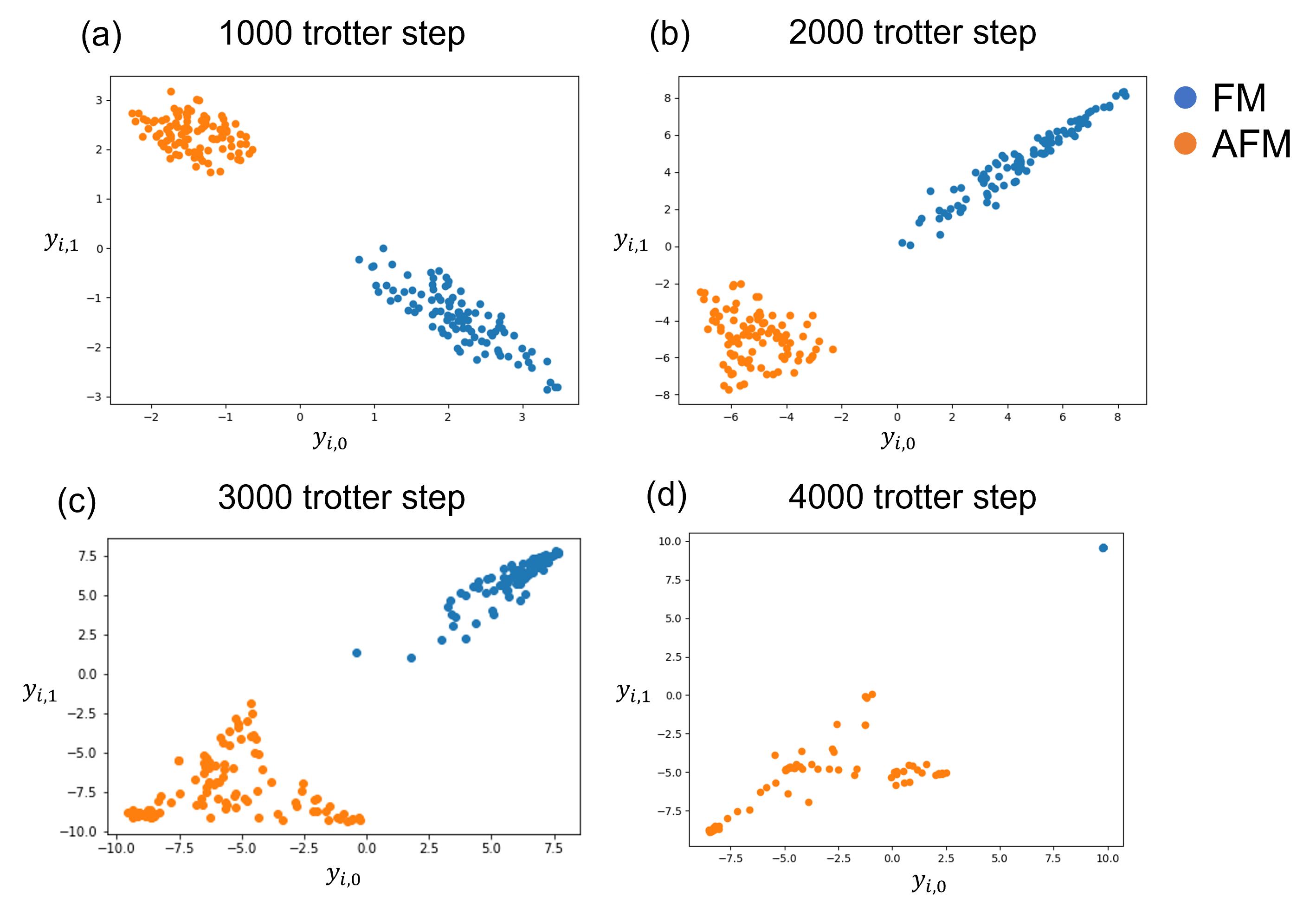}
    \caption{
    These figures illustrate the process of a time-dependent transverse-field two-body Ising model when the calculation of the similarity in high-dimensional data is based on the fidelity. 
    The initial state is $|+\rangle^{\otimes 4}$.
    The upper left and upper right figures visualize the quantum states of the $1000$th and $2000$th trotter step, respectively. 
    The lower left and lower right figures show the $3000$th trotter step and the last $4000$th trotter step, respectively.
    }
    \label{fig:qa_hdist_dot_output}
\end{figure*}

The second method is to calculate the similarity of the data in high-dimensional space 
via infidelity of two different quantum states.
The similarity between low-dimensional data
is calculated in the same way as the previous case,
except for the setting of hyper-parameter $a=10$.
Fig.~\ref{fig:qa_hdist_dot_output} shows 
the process of the dynamics for every $1000$ trotter steps. 
This figure also shows the two clusters corresponding to the sign of the coupling constants.

\subsubsection{The effect of multiplying a constant value}
\begin{figure*}
    \centering
    \includegraphics[clip, keepaspectratio, scale=0.5]{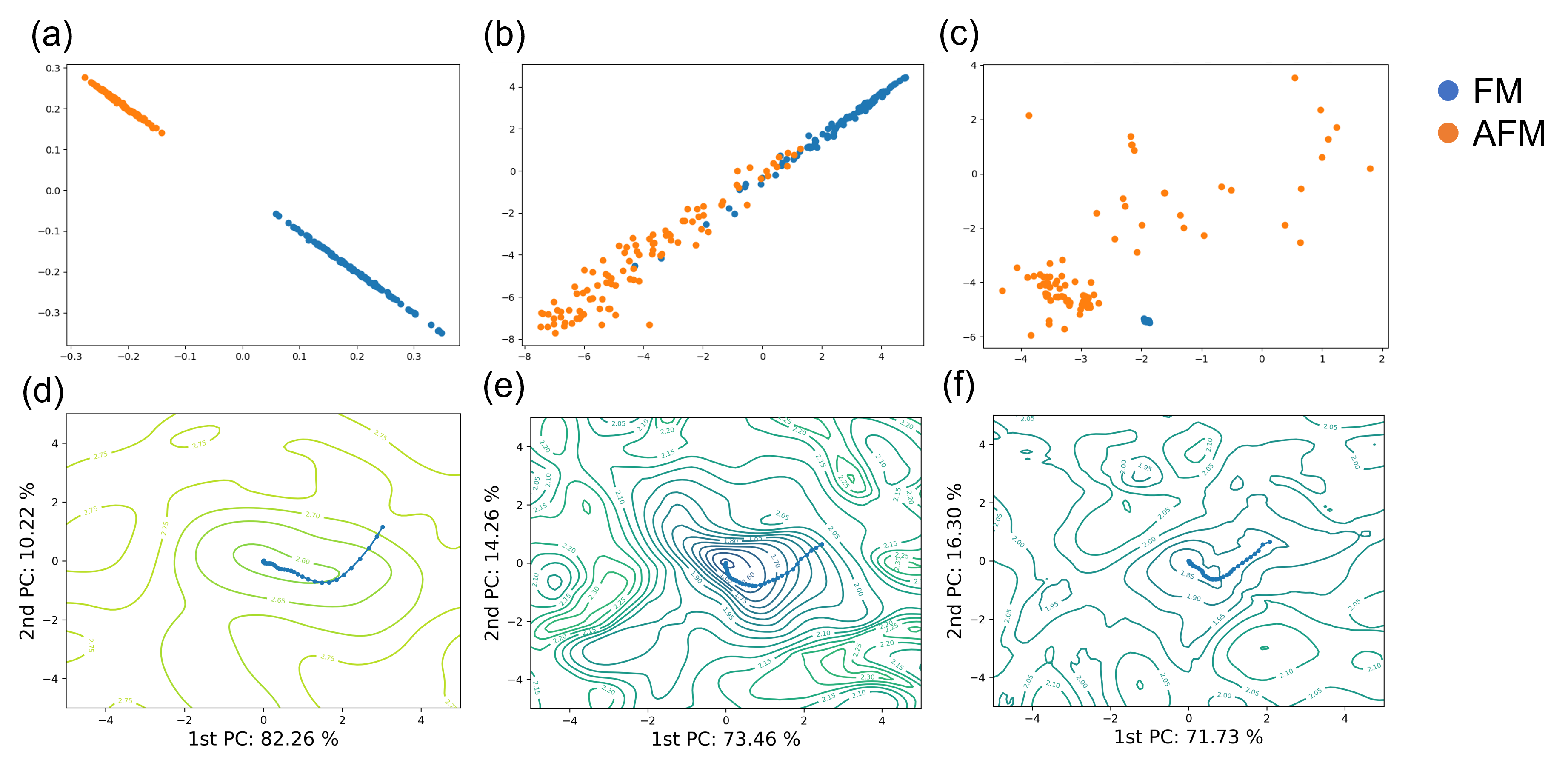}
    \caption{
    These are examples of undesirable visualizations. 
    In Fig.~(a), although the plot is two-dimensional, 
    it is plotted in a one-dimensional manner. 
    In Figs.~(b) and (c), the different classes cannot be identified unless they are plotted in different colors.
    Figs.~(d), (e), and (f) are the loss contours and optimization trajectories of Fig.~(a), (b), and (c), respectively.
    In all three cases, the parameters are optimized.
    }
    \label{fig:vis_fail}
\end{figure*}

When we visualize the quantum data, 
we sometimes get undesirable visualization figures,
which can be avoided by appropriately tuning the hyperparameter $a$.
For example, we plot three examples of wrong choices of $a$ in Fig.~\ref{fig:vis_fail}.
In Fig.~\ref{fig:vis_fail} (a), the data is arranged in an almost straight line,
despite the two-dimensional plot. 
In figures (b) and (c), 
it is impossible to determine that these data belong to different classes 
without the colors in the plots, 
since each class is colored to make the figures easier to understand.
One possible reason is that the cost function optimization is not sufficient to converge to a local minimum.

We examine whether the optimization converges near a local minimum.
To this end, we visualize the loss contours and optimization trajectories in two dimensional plane.
In Ref.~\cite{li2017visualizing}, 
they develop a method to visualize loss landscape and optimization trajectories with the loss contours.
They use the visualization method to investigate the relationship 
between loss landscape and trainability or generalization for NNs.
Before we describe the detail method, 
let us define $\theta_i$ as a vector of trainable parameters at $i$th epoch.
We consider the matrix $M$ consisting of $(\theta_i - \theta_n)$ for all $i$, 
where the $n$ is the last epoch. 
Specifically, the $M$ is written by
\begin{eqnarray} \label{eqn:vis_matrix}
M = [\theta_0-\theta_n;\theta_1-\theta_n;\ldots;\theta_{n-1}-\theta_n]. 
\end{eqnarray}
The matrix $M$ is transformed by principal component analysis \cite{pearson1901liii},
and the $1$st and $2$nd principal component vectors are used as the axis of the visualization.
Along the two principal component vectors, the loss contours and optimization trajectories are plotted on two-dimensional plane.
We use this visualization method to tune the hyperparameter $a$ and to confirm our optimization works appropriately.

We show the visualization of the loss landscapes in Fig.~\ref{fig:vis_fail} (d), (e), and (f).
From these figures, the optimization is sufficiently done to reach the local minima.
Despite the cost function being well optimized,
we have not been able to reflect enough features on the low-dimensional data to discriminate different clusters.
This is attributed to the wrong design of the cost function, 
where the similarity is not appropriately defined.
For example, in Fig.~\ref{fig:vis_fail} (a), 
the parameter $a$ for adjusting the scale is multiplied by $1$ instead of by $10$ in Fig.~\ref{fig:qa_hdist_dot_output} (a). 
In Fig.~\ref{fig:vis_fail} (b) and (c), both high-dimensional and low-dimensional data are multiplied by $10$ in similarity calculation, instead of by $1$ in Fig.~\ref{fig:qa_output} (c) and (d).
These results suggests that 
when calculating similarity it is necessary to adjust the scale of the high- and low- dimensional data 
so that the cost function adequately reflects the characteristics of the given data.

\section{\label{sec:conclusion}Conclusion and Future works}
In this work, 
we have visualized classical and quantum data by quantum neural network. 
Specifically, we employed a parameterized quantum circuit 
as a quantum model to generate low-dimensional data,
which is trained so that the similarity of the high-dimensional data 
is maintained.
In the quantum case, the similarity of the quantum states 
is calculated in two ways: measuring multiple observables to obtain high-dimensional classical data from quantum states, and calculating distance of two quantum states directly from fidelity.

We performed numerical simulation of two-dimensional visualization of Iris dataset as the case of the classical inputs and quantum states evolved under Hamiltonian dynamics as the case of the quantum inputs.
It was found that the proposed method worked well for both classical and quantum inputs, with appropriate low-dimensional visualization.
Specifically, 
in the case of quantum data, it is difficult to visualize the data well unless the constant factor is multiplied on the similarity of the low-dimensional data.
This is probably because the similarity between the quantum states takes too small value. While we treated this constant factor as a hyperparameter, it can be trainable parameters as $a_i$ for each low-dimensional data $y_i$ for further improvement. At least, by doing so, the performance of non-parametric t-SNE can be guaranteed in the proposed model with a two-qubit trivial circuit. This implies that the proposed method is expected to work on a relatively small quantum device.
For more complex data visualization, we can improve the representation capability of quantum circuits.

One of the other possibilities using our proposed method is to compress quantum data 
by defining the low-dimensional data by quantum states and the similarity by a fidelity-based metric.
While real quantum data, such as a set of outputs from a quantum algorithm, live in a large Hilbert space,
such quantum data can be mapped into a smaller Hilbert space with a fewer number of qubits with keeping their similarity.
Then quantum machine learning algorithms can be further applied on such a compressed quantum data set.
Further investigation in this direction is an intriguing future issue.
The other future task is to construct a model including a decoder, which decodes data from compressed data,
such as Variational AutoEncoder \cite{kingma2013auto} in classical machine learning.
It may be possible to create a quantum state with certain desired properties from the classical data in the middle layer.

\begin{acknowledgments}
K.M. is supported by JST PRESTO Grant No. JPMJPR2019 and JSPS KAKENHI Grant No. 20K22330.
K.F. is supported by JST ERATO Grant No. JPMJER1601 and JST CREST Grant No. JPMJCR1673.
This work is supported by MEXT Quantum Leap Flagship Program (MEXTQLEAP) Grant No. JPMXS0118067394 and JPMXS0120319794.
We also acknowledge support from JSTCOI-NEXT program.
\end{acknowledgments}

\appendix

\bibliography{apssamp}

\end{document}